\documentclass[9pt,twocolumn,twoside]{osajnl}

\journal{ol} 

\setboolean{shortarticle}{true}

\title{Single-Pixel Pattern Recognition with Coherent Nonlinear Optics}

\author[1,2]{Ting Bu}
\author[1,2]{Santosh Kumar}
\author[1,2]{He Zhang}
\author[1]{Irwin Huang}
\author[1,2,3*]{Yuping Huang}

\affil[1]{Department of Physics, Stevens Institute of Technology, Hoboken, NJ, 07030, USA}
\affil[2]{Center for Quantum Science and Engineering, Stevens Institute of Technology, Hoboken, NJ, 07030, USA}
\affil[3]{Department of Electrical and Computer Engineering, Stevens Institute of Technology, Hoboken, NJ, 07030, USA}
\affil[*]{Corresponding author: yhuang5@stevens.edu}

\begin{abstract}
We propose and experimentally demonstrate a nonlinear-optics approach to pattern recognition with single-pixel imaging and deep neural network. It employs mode selective image up-conversion to project a raw image onto a set of coherent spatial modes, whereby its signature features are extracted nonlinear-optically. With 40 projection modes, the classification accuracy reaches a high value of 99.49\% for the MNIST handwritten digit images, and up to 95.32\% even when they are mixed with strong noise. Our experiment harnesses rich coherent processes in nonlinear optics for efficient machine learning, with potential applications in online classification of large-size images, fast lidar data analyses, complex pattern recognition, and so on. 

\end{abstract}

\setboolean{displaycopyright}{true}

\begin{document}

\maketitle


Artificial intelligence is a powerful tool to perform and expedite prohibitively complex tasks such as image classification \cite{Rahmani2020}, speech recognition \cite{deng2013new}, drug discovery and genetics \cite{Chen2016}. Among many exciting areas, machine learning for recognizing patterns or images with high speed and accuracy has become a subject of focused interest due to its many applications in healthcare \cite{Deep-STORM,Rivenson:18}, robotics \cite{Palagi2016,Wani2017}, metrology \cite{VanderJeught:19,Qian:20}, information processing \cite{steinbrecher_quantum_2019}, etc. For high-resolution image reconstruction and classification, deep learning---a branch of machine learning---has achieved considerable successes and demonstrated great potential \cite{lecun2015d, rivenson2017deep,bulgarevich_pattern_2018,Metzler:20}. By utilizing a feed-forward multi-layer artificial neural network, its  performance has been shown to be superior in certain complex imaging tasks \cite{zahavy_deep_2018,ziv_deep_2020,Wengrowicz:20,PRL_photonphaseRet}. 

Recently, optical neural networks have emerged as distinct candidates for solving complex problems, by leveraging their large data parallelism, high connectivity, low power consumption, and other advantages inherent with optical circuits and architectures  \cite{1998_Yu,Zuo:19,PRX_Hamerly,zhou2020situ}. In this pursuit, optical Fourier transformation, diffraction, interference and spatial filtering have been utilized for optical pattern restoration and recognition \cite{lin2018a,PourFard:20,PRL_FDNN}, phase retrieval, and information processing \cite{cheng_optical_2019,Liu:19}. While most developments use only linear optics \cite{kleinert2019r,PRL_Liu2019,PRL_Giordani2020,zhai2020t,Hossein2020}, nonlinear optical effects have been shown to assist well phase retrieval of ultrashort pulses \cite{zahavy_deep_2018,ziv_deep_2020}, polaritonic neuromorphic computing \cite{ballarini2020p}, and the classification of ordered and disordered phases of Ising model \cite{zuo_all-optical_2019}. As nonlinear optics can realize even richer and more complex functions than its linear counterparts, nonlinear optical machine learning (NOML) promises another level of data processing capability and efficiency.   

In this letter, we demonstrate a new nonlinear optics paradigm for efficient and robust machine learning. It is based on mode-selective image conversion (MSIC) by spatially modulated sum-frequency 
generation, an exceptional nonlinear optical technique that we recently demonstrated for photon efficient classification with super-resolution \cite{Zhang:20} and mode selective detection through turbulent media \cite{Santosh19,zhang_mode_2019}. MSIC is implemented by applying a spatially structured pump beam to drive the frequency conversion in a nonlinear crystal, so that only a single signal mode of certain prescribed phase coherence can be converted efficiently. All other modes, even if they spatially, temporally, and spectrally overlap with the signal mode, are not converted or converted with a much lower efficiency. It thus realizes a distinct tool to sort optical spatial modes according to their phase coherence signatures. Unlike linear optical approaches, MSIC does not involve any direct modulation on the signal, thus eliminating the otherwise inevitable modulation loss or noise \cite{koprulu2011l}. Also, it allows much more flexible and capable operations by engineering the nonlinear dynamics in the crystal with the assistance of spatial dispersion  \cite{zhang_mode_2019}.

\begin{figure*}[ht]
\centering
\includegraphics[width=0.9\linewidth]{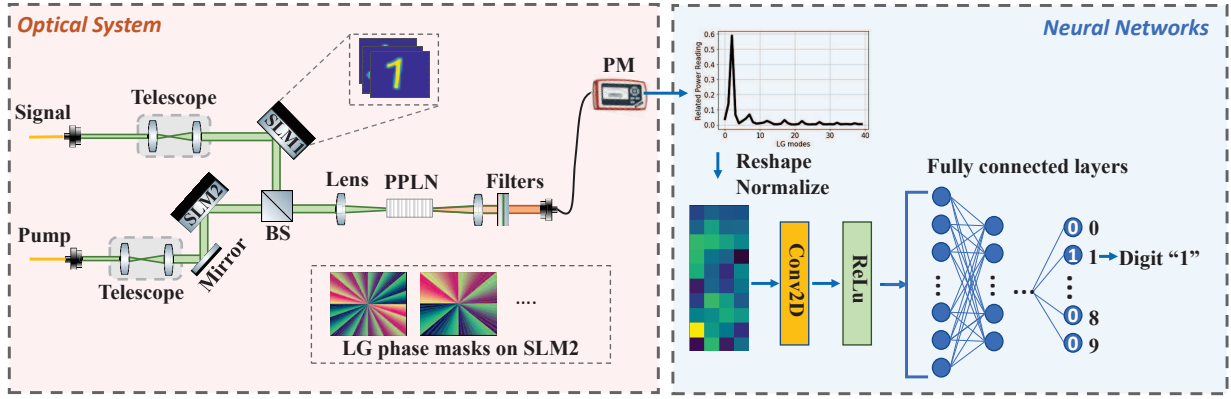}
\caption{The system setup with nonlinear optics and DNN parts. The signal and pump beams, each at wavelength 1545 and 1558 nm, are generated from a common mode lock laser. The spatial light modulator (SLM1) on the signal arm is used to upload the phase patterns of handwritten digits and the other (SLM2) on the pump arm is used to create the LG modes for the feature extraction. These two beams are then combined using a beamsplitter (BS) and passed through a temperature stabilized PPLN crystal. The generated SF light at 775.5 nm is filtered and coupled into a single mode fiber. Its optical power is read by an optical power meter (PM) and sent to the computer. The neural networks consisting of one convolution layer and fully connected layers are used for the training of digits recognition. BS: Beamsplitter, SLM: Spatial Light Modulator, PPLN: Magnesium-doped Periodic Poled Lithium Niobate crystal, Conv 2D: Two-dimensional convolution layer and ReLu: Rectified Linear Unit.}
\label{fig:setup}
\end{figure*}

Here, we incorporate MSIC and a deep neural network (DNN) for pattern recognition of high-resolution images. In this hybrid architecture, the information in each input image is first processed and extracted by MSIC, through frequency up-conversion by pump beams in 40 different Laguerre-Gaussian (LG) modes. For each mode, the converted power is measured and used as the input to the DNN for subsequent machine learning. In this design, MSIC pre-processes the input images, extracting information contained in both amplitude and phase spatial profiles of the images, and condense them into the upconversion efficiencies of 40 modes. It reduces a large amount of pixel-wise information to only 40 signatures, thus substantially downsizing the required DNN while enabling efficient processing of high resolution images. As a benchmark, our experiment achieves an accuracy of 99.49\% for recognizing the MNIST (Modified National Institute of Standards and Technology) handwritten digit images, in a good agreement with our theoretical value of 99.83\%. Even when strong noise is added to each image, whose resulting mean signal-to-noise ratio (SNR) is about $-11.2$ dB for the image, an accuracy of 95.32\% can still be reached. 

The experimental setup is shown in Fig.~\ref{fig:setup}. A mode-lock laser generates optical pulses with $\sim$300 fs full width at half maximum (FWHM) and 50 MHz repetition rate. Two inline narrow-band  wavelength division multiplexers (WDMs, bandwidth $\sim$ 0.8 nm) select two separate wavelengths, one at 1558 nm as the pump and the other at 1545 nm as the signal. The two pulse trains are then each amplified using an Erbium-doped fiber amplifier. They are then aligned to be in horizontal polarization using free-space optics and expanded to a 2.8-mm beam waist (FWHM) for the pump and 2.6 mm for the signal. Then, the signal is directed to a reflective, liquid-crystal spatial light modulator (SLM1 in Fig.~\ref{fig:setup}) at a $55^{\circ}$ incidence angle, while the pump to another modulator of the same model (SLM2 in Fig.~\ref{fig:setup}) but at a $50^{\circ}$ angle \cite{Santosh19}. The SLMs are Santec SLM-100 and have the same specs: pixel pitch $\sim$ 10.4 $\times$ 10.4 $\mu$m, active area $\sim$ 1.4 $\times$ 1 cm. 

In our experiment, the MNIST digit images are uploaded onto SLM1 as phase mask patterns. To prepare the pump in a sequence of LG modes, their helical phase patterns are expressed on SLM2 in the form of $\Theta(r,\phi) = -l\phi +\pi \theta (-L_p^{|l|}(2r^2/w_0^2))$, where $\Theta$ is wrapped between 0 and $2\pi$,  $\theta$ is a unit step function, $\{\displaystyle L_{p}^{\vert l \vert}\}$ are the generalized Laguerre polynomials with the azimuthal mode index $l$ and the radial index $p$, $w_0$ is the beam waist, and $\phi=\arctan(y/x)$ is the azimuthal  coordinate (see \cite{zhang_mode_2019} for more information). The SLM response time is typically 500 ms, which varies with the structural complexity of the phase patterns. The pump and signal beams are then merged at a beam splitter (BS) and focused ($f$=200 mm) inside a temperature-stabilized periodic poled lithium niobate (PPLN) crystal with poling period of 19.36 $\mu$m and the total length of 1 cm (5 mol.\% MgO doped PPLN, from HC Photonics) for MSIC. The normalized conversion efficiency of the crystal, assuming optimal focusing inside the crystal, is $\sim$ 1\%/W/cm. The generated SF light is coupled into a single-mode fiber and detected by a power meter (Thorlabs, PM-100D with sensor S130C) through a MATLAB interface.

The SF power readings are fed into the DNN after normalization (see the following paragraphs), which consists of one convolutional layer, five fully connected layers, and one output layer with 10 neurons for the 10 different classes, as shown in the right panel of Fig. \ref{fig:setup}. There are 16 filters in the convolutional layer with kernel size of 2, and the five fully connected layers have 512, 256, 128, 64, and 32 units, respectively.

The nonlinear activation function rectified linear unit [ReLu$(x)=max(0,x)$] is used in each connection between hidden layers because of its faster convergence than other nonlinear functions \cite{wang2019effi,kri2012imagenet}. The categorical cross entropy is selected as the Loss function to evaluate the performance of the DNN 
\begin{equation}
\textrm{Loss}=-\sum_{k=1}^{K}\sum_{i=1}^{N}y_{ik}log(\hat{y}_{ik}),\label{eq:loss}
\end{equation}
where $y_{ik}$ is the true output, $\hat{y}_{ik}$ is the predicted output, $N$ is the total number of samples, and $K=10$ is the number of classes. To minimize the Loss function, adaptive moment estimation (ADAM) gradient descent algorithm is used as the optimizer in the training process \cite{kingma2014adam}. Then the softmax function is adopted to normalize the values in output layer so that they represent the probability of each class, as 
$P(j)=\frac{e^{x_{j}}}{\sum _{k=1}^{K}e^{x_{k}}}$,
where $j=1,...,K$ 
and $x$ is the probability vector without normalization. Finally, 
the predicted classification for each input will be picked as the one with the highest probability, i.e., the ``winner'' of all classes, and the accuracy score of the whole database is the fraction of correctly classified samples.

To determine the optimal set of the pump LG modes, we start from simulating the performance with 110 LG modes with $l\in[-5,5]$ and $p\in[0,9]$, as the SF generation efficiency is negligible for other modes. Then, based on the simulated  classification accuracy for different combination of LG modes, we choose a subset of most effective 40 modes with $l\in[-2,2]$ and $p\in[0,7]$ as the pump modes for our experiments.

For the benchmark evaluation, we select the first 600 handwritten images of each digit in the MNIST database as our dataset. The resolution of those images are first increased by the Image Resize function in Matlab from 28$\times$28 to 400$\times$400 pixels to match with the SLM pitch and signal beam diameter. 
After collecting SF power readings for all images, the data is shuffled and separated into training (4800 images) and testing sets (1200 images). With the training set, for each LG mode of the pump, the power readings are normalized across all digits so that different features extracted by those modes contribute more or less equally to the machine learning. Afterwards, the 40 normalized readings for each image are mapped onto a 4$\times$10 matrix for feeding into the DNN for training on a NVIDIA Quadro T2000 graphics processing unit (GPU). 

\begin{figure}[ht]
\centering
\includegraphics[width=1\linewidth]{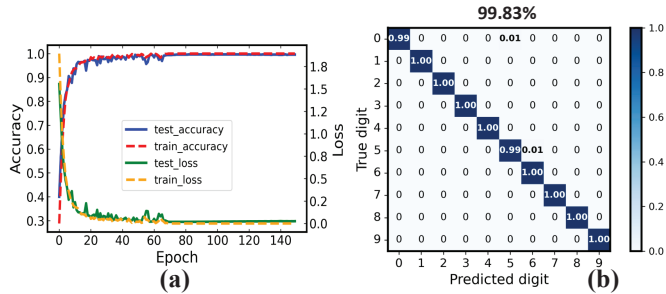}
\caption{Performance of simulated results. (a): Classification accuracy as a function of epoch for training and testing database. (b): Normalized confusion matrix for testing database.}
\label{fig:results_wonoise}
\end{figure}

Prior to the experiments, we first perform some simulation studies to understand the nonlinear optical process and overall system behavior. Figure \ref{fig:results_wonoise}(a) plots the simulated accuracy as a function of the epoch number during learning process. It shows that the accuracy scores for both training and testing dataset converge stably to high values. At the end, the recognition accuracy of 100\% is achieved for the training and 99.83\% for the testing dataset. Figure \ref{fig:results_wonoise}(b) presents the final normalized confusion matrix for the testing set, with nearly perfect classification for all digits. This indicates that our model is well trained without over-fitting or under-fitting. 

In the current experiment, the overall system speed is limited by the SLM response time ($\sim$ 0.5 s), due to which a pause time of 1.5 s is set in the experiments to ensure a well uploaded mask. Therefore, it takes days to measure all MNIST images. While the system is stable during short time, over the span of days there are significant mechanical fluctuations, beam drifts, and ambient noise in our table-top settings over free space. All lead to measurement errors and biases that reduce the recognition accuracy. To access those effects, we adopt two data collection procedures. The first uses a sequential run, where the 600 images for each digit are measured in a group before moving to the next digit. The second runs in a loop, with the images measured in groups of 10, with each containing one image for every digit. 

\begin{figure}[ht]
\centering
\includegraphics[width=1\linewidth]{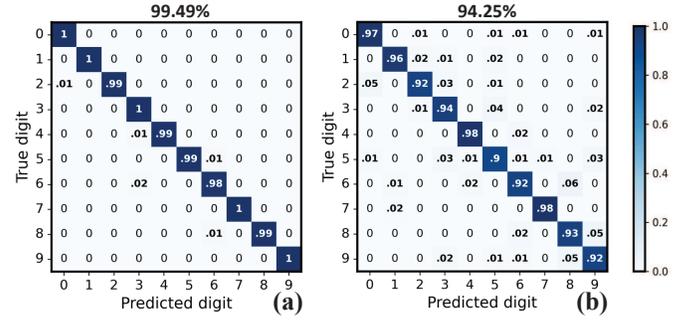}
\caption{Normalized confusion matrix for experimental results with (a) sequential run and (b) loop run for the original database.}
\label{fig:train_acc} 
\end{figure}

Figure \ref{fig:train_acc} presents normalized confusion matrix of the recognition accuracy for each digit. The sequential run reaches an accuracy score of 99.49\%, which approaches to the theoretical value of 99.83\% obtained in our simulation. Such an excellent agreement validates our method and the underlying optics. With the loop run, on the other hand, the accuracy is dropped to 94.25\%. This may be caused by the experimental instabilities and noise during the elongated data taking, so that the signature features for each digit are disrupted and blurred. In the future, this issue can be overcome by designing a more compact and enclosed system. Also, the low-speed SLMs can be replaced with fast digital micromirror devices (DMDs) to substantially shorten the data taking cycle.

 \begin{figure}[ht]
\centering
\includegraphics[width=0.8\linewidth]{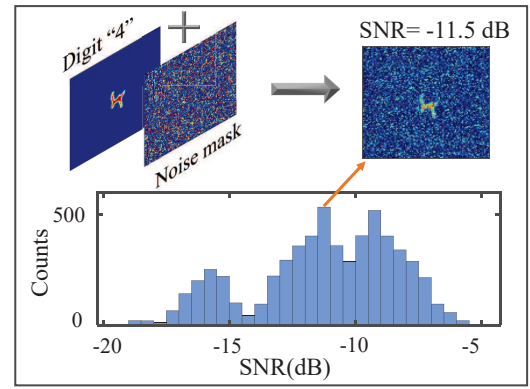}
\caption{The histogram of SNR distribution of the database and an example of noise-added digit "4" with SNR=-11.5 dB.}
\label{fig:noise_image} 
\end{figure}
 
To further benchmark the present hybrid pattern recognition technique, we next test the classification of the handwritten images with added random noise. Figure \ref{fig:noise_image} shows an example of such random noise as a phase mask added to a resized MNIST image. Each noise spot in the noise mask is in a cluster with 10$\times$10 pixels, which is designed as stronger noise than smaller clusters. The noise cluster each have random phase values uniformly distributed between $[0,2\pi)$, and together they take up 40\% of the total pixels. To quantify the resulting signal to noise, we define a SNR in decibel as
\begin{equation}
    \textrm{SNR}=10\times log_{10}(\sigma _{s}^{2}/\sigma _{n}^{2}),
\label{eq:snr}
\end{equation}
where $\sigma _{s}^{2}$ and $\sigma _{n}^{2}$ are the phase variance of a digit image and noise mask, respectively.  
As shown in Fig.~\ref{fig:noise_image}, the SNR in the current experiment varies between -19.2 dB and -5.3 dB, with a mean value -11.2 dB.
 
\begin{figure}[ht]
\centering
\includegraphics[width=1\linewidth]{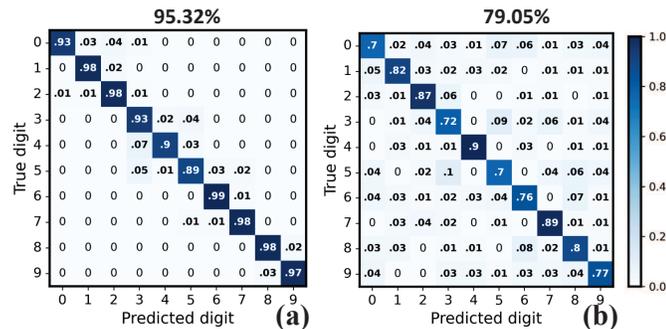}
\caption{Normalized confusion matrix for the noisy images with (a) sequential and (b) loop experiment runs.}
\label{fig:results_wnoise} 
\end{figure}

Figure~\ref{fig:results_wnoise} presents the normalized confusion matrices of the MNIST database with added noise. By the sequential run, the accuracy reaches 95.32\%, with the low accuracy values all locating near the diagonal line. It indicates that the noisy digit images are prone to be mis-classified with its neighboring digits. 

By contrast, the recognition accuracy is only 79.05\% in the loop run case, marking an enlarged drop than the non-noise case as in Fig.~\ref{fig:train_acc}. This indicates that the noisy digit images become more prone to the experimental instabilities and ambient interference, as the signature features are blurred by the added noise. 

In summary, we have demonstrated a hybrid machine learning system incorporating nonlinear optics and a deep neural network. It uses mode-selective image conversion--realized through a coherent nonlinear optical process in a $\chi^{(2)}$ crystal--to extract the signature features of images for subsequent machine learning. The nonlinear optics step utilizes all-to-all connection and speed-of-light processing of large-volume image data over hundreds of thousands, which are otherwise overwhelming for typical electronic digital processors. Using 40 pump modes for the information extraction, we demonstrate in experiment handwritten digital classification at a high accuracy of 99.49\%, close to the theoretical result of 99.83\%. Even when the images are mixed with significant noise to have a mean signal to noise ratio of -11.2 dB, the classification accuracy can still exceed 95\%. 
Our results indicate the viability and potential advantages of introducing coherent nonlinear optics to machine learning and artificial intelligence. For the future work, we hope to replace the current SLMs with much higher speed devices such as those of micro-mirrors, where orders of magnitude speedup is expected. Also, the current setup faces the challenges due to free-space optics fluctuations and ambient noises, which shall be addressed by an improved, enclosed design.


\noindent\textbf{Disclosures.} The authors declare no conflicts of interest.



\end{document}